\documentclass[twocolumn,showpacs,preprintnumbers,amsmath,amssymb]{revtex4}

\usepackage{graphicx}
\usepackage{dcolumn}
\usepackage{epsfig}
\usepackage{bm}
\usepackage{draftcopy}

\begin{document}

\preprint{}

\title{Generalized Parton Distributions from Deeply Virtual Compton 
Scattering at HERMES}

\author{M. Guidal}
\email{guidal@ipno.in2p3.fr}
\affiliation{%
Institut de Physique Nucl\'eaire d'Orsay,\\
91405 Orsay, FRANCE
}%

\author{H. Moutarde}
\email{herve.moutarde@cea.fr}
\affiliation{
CEA Saclay\\
Service de Physique Nucl\'eaire \\
91191 Gif-Sur-Yvette, France}

\date{\today}

\begin{abstract}
The HERMES collaboration has recently published a set of 
(correlated) beam charge, beam spin and target spin asymmetries
for the Deeply Virtual Compton Scattering process. This reaction allows in
principle to access the Generalized Parton Distributions of the nucleon.
We have fitted, in the QCD leading-order and leading-twist handbag 
approximation, but in a model independent way, this set of data and 
we report our results for the extracted Compton Form Factors. In particular,
we are able to extract constrains on the $H$ GPD.
\end{abstract}


\maketitle

More than forty years after the discovery of partons inside the nucleon, 
the precise way they compose the nucleon and give rise to its
properties remains a large mystery. In this past decade,
the powerful concept of Generalized Parton Distributions (GPDs), which allows 
to describe the nucleon structure in an unprecedented way, has emerged. Among other features, 
the GPDs contain the correlations between the spatial and 
momentum distributions of the partons inside the nucleon. These correlations,
which are currently unknown, can thus allow for a tomographic image of the 
nucleon, by providing spatial distributions of the quarks for different 
momentum ``slices". Besides such an imaging, measuring GPDs allows to access some 
fundamental properties of the nucleon such as, through Ji's sum rule, 
the orbital momentum contribution of the quarks to the spin of the nucleon
(which is, classically, the cross product of position and momentum).
More generally, GPDs reflect the complex non-perturbative
dynamics of the strong force which governs the interaction between
quarks and gluons. It is one of the main challenges of today's modern
physics to understand the confinement regime of the strong force and
its associated theory, Quantum Chromo-Dynamics (QCD). Measuring
GPDs, which can be modeled or calculated theoretically, certainly opens 
a wide new window onto these fundamental issues.

\begin{figure}[htb]
\epsfxsize=9.cm
\epsfysize=10.cm
\epsffile{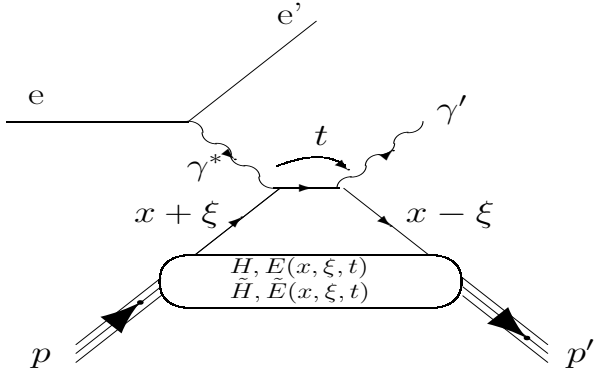}
\vspace{-4.2cm}
\caption{The handbag diagram for the DVCS process on the proton $ep\to e'p'\gamma '$. 
Here $x+\xi$ and $x-\xi$ are the longitudinal momentum fractions of the 
initial and final quark, respectively, and $t=(p-p')^2$ is the squared momentum 
transfer between the initial and final protons (or equivalently between the two
photons). There is also a crossed diagram which is not shown here.}
\label{fig:dvcs}
\end{figure}

We refer the reader to refs.~\cite{muller,ji,rady,collins,goeke,revdiehl,
revrady,barbara} for the original theoretical articles and recent comprehensive 
reviews on GPDs and for more details on the theoretical formalism.
In short, formally, the GPDs are the Fourier transforms of 
matrix elements of QCD light-cone 
bilocal operators between nucleon states of different momenta.
They can be measured through Deeply Virtual 
Exclusive Processes such as, for the process the simplest theoretically and the most 
accessible experimentally, the exclusive leptoproduction of a photon or a 
meson on a nucleon at large $Q^2$ (the lepton's squared momentum-transfer). 
In the following, we focus on the Deeply Virtual 
Compton Scattering (DVCS) process on the proton, $ep\to ep\gamma$, 
which has been recently the subject of an intense experimental 
effort~\cite{franck,fx,ave,zeiler}. The process is illustrated on 
fig.~\ref{fig:dvcs}. 

The amplitude of
the DVCS process is the convolution of the GPDs and a hard kernel which 
represents the Compton scattering of a (virtual) photon with a quark
of the nucleon, which can be calculated perturbatively in both
QED (Quantum Electro-Dynamics) and QCD. For helicity conserving quantities in the 
quark sector, there are four GPDs, $H, \tilde H, E, \tilde E$ which depend, 
in leading order QCD perturbation theory and leading twist QCD, on three 
variables: $x$, $\xi$ and 
$t$. Both $x$ and $\xi$ express the longitudinal momentum fractions of the 
two quarks of the bilocal operator, while $t$ is the squared four-momentum 
transfer between the final and initial nucleon. $\xi$ is related to the 
well-known Deep Inelastic Scattering (DIS) variable $x_B$: 
$\xi\approx\frac{x_B}{2-x_B}$ (at leading order).

In processes such as DVCS, if the $\xi$ and $t$ variables are well measurable
experimentally (by measuring the kinematics, respectively, of the scattered
lepton and the recoil nucleon -or the final state photon-), the $x$ variable is
a ``mute" variable. Indeed, the GPDs enter the DVCS amplitude under the form 
of a convolution integral over $x$:
$\int_{-1}^{+1}d x {{GPD(x,\xi,t)} \over {x + \xi - i \epsilon}}$.
Decomposing this expression into a real and an imaginary part
(and reducing the $x$-range from $\{-1,1\}$ to $\{0,1\}$), there are
therefore in principle eight GPD-related quantities that can be extracted in the DVCS
process, in the QCD leading twist and leading order approximation, which is the frame of this study:
\begin{eqnarray}
&&H_{Re}=P \int_0^1 d x \left[ H(x, \xi, t) - H(-x, \xi, t) \right] C^+(x, \xi),\label{eq:eighta} 
\\
&&E_{Re}=P \int_0^1 d x \left[ E(x, \xi, t) - E(-x, \xi, t) \right] C^+(x, \xi),\label{eq:eightb} 
\\
&&\tilde{H}_{Re}=P \int_0^1 d x \left[ \tilde H(x, \xi, t) + \tilde H(-x, \xi, t) \right] C^-(x,
\xi),\label{eq:eightc} 
\\
&&\tilde{E}_{Re}=P \int_0^1 d x \left[ \tilde E(x, \xi, t) + \tilde E(-x, \xi, t) \right] C^-(x,
\xi),\label{eq:eightd} 
\\
&& H_{Im}=H(\xi , \xi, t) - H(- \xi, \xi, t),\label{eq:eighte} \\
&& E_{Im}=E(\xi , \xi, t) - E(- \xi, \xi, t),\label{eq:eightf} \\
&& \tilde{H}_{Im}=\tilde H(\xi , \xi, t) + \tilde H(- \xi, \xi, t) \;\;\;\;\text{and}\label{eq:eightg} \\
&& \tilde{E}_{Im}=\tilde E(\xi , \xi, t) + \tilde E(- \xi, \xi, t)\label{eq:eighth} 
\end{eqnarray}

with 
\begin{equation}
C^\pm(x, \xi) = \frac{1}{x - \xi} \pm \frac{1}{x + \xi}.
\end{equation}

In order to avoid any potential confusion for the $Re$ and $Im$ symbols, let us note that 
we have slightly changed our notation with respect to ref.~\cite{fitmick} 
where these eight quantities were actually called, respectively, $Re(H)$, $Re(E)$, $Re(\tilde{H})$, 
$Re(\tilde{E})$, $Im(H)$, $Im(E)$, $Im(\tilde{H})$ and $Im(\tilde{E})$. We will call 
them the Compton Form Factors (CFFs), in a slight abuse of terminology. We point
out that in our definition there is a $\pi$ factor for the $H_{Im}$, $E_{Im}$, $\tilde{H}_{Im}$ and 
$\tilde{E}_{Im}$ CFFs and a ``-" sign for the $H_{Re}$, $E_{Re}$, $\tilde{H}_{Re}$, 
$\tilde{E}_{Re}$ CFFs of difference with respect to the original CFF definition of ref.~\cite{kirch}.
Since we are focusing in this study on the proton, it has to be understood in the following 
that we consider proton GPDs, $i.e.$ 
$GPD(x , \xi, t)=\frac{4}{9}GPD^u(x , \xi, t)+\frac{1}{9}GPD^d(x , \xi, t)$.

Also, we recall that, experimentally, there is another process leading to the
same final state $ep\to ep\gamma$ than the DVCS: the Bethe-Heitler (BH) process, 
in which the final state photon is radiated by the incoming or scattered electron and not 
by the nucleon itself. The BH process is very well calculable in QED and the
only non-perturbative QCD quantities entering are the proton form factors, which are 
relatively precisely known at the relevant kinematics, $i.e.$ small $t$.

Now the question arises: given the well calculable leading-twist DVCS and BH amplitudes,
to which extent the CFFs, which are unknown quantities, can be extracted from real data ? 
The problem is not trivial since the formulas linking the CFFs to the observables
are complex and non-linear. A dedicated study based 
on Monte-Carlo simulations has been carried out in ref.~\cite{fitmick} showing that,
given enough observables (charge-~, single and double-spin observables) and enough 
experimental accuracy (of the order of a few percent), essentially all CFFs could be 
extracted from (pseudo-)data.
This study was based on the combination of the well-known CERN minimization program 
MINUIT~\cite{james} and the VGG code~\cite{vgg1,gprv} which calculates the DVCS 
and BH amplitudes and the associated observables. The VGG code also provides parametrizations 
of the GPDs. As a real application, the recent Jefferson Laboratory (JLab) Hall A unpolarized 
and beam-polarized cross sections data~\cite{franck} were fitted, this resulting in the 
first ever 
model-independent (leading-twist) extraction of the $H_{Re}$ and $H_{Im}$ CFFs, though with 
large uncertainties since only two observables were fitted.

In the present study, our aim is now to apply our fitting procedure to the 
recent set of DVCS asymmetries recently published by the HERMES collaboration. 
In short, compared to the JLab/Hall A experiment, the HERMES experiment,
although in a different kinematic regime, provides more observables but with 
lesser precision. The question is then: can we still extract some (model independent) 
GPD information ? 

\begin{figure}[htb]
\epsfxsize=10.cm
\epsfysize=10.cm
\epsffile{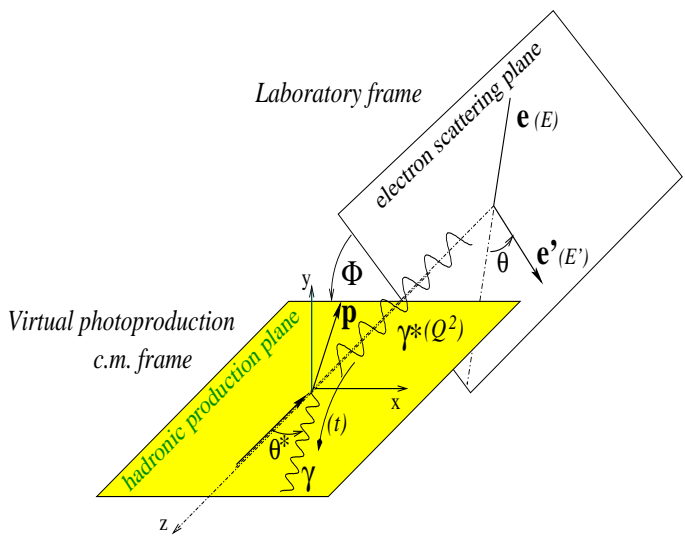}
\vspace{-2.5cm}
\caption{Reference frame and relevant variables for the description of the 
$ep\to ep\gamma$ reaction.}
\label{fig:frame}
\end{figure}

Precisely, the HERMES collaboration has measured~\cite{ave,zeiler}:
\begin{itemize}
\item charge asymmetries~\cite{ave,zeiler}:
\begin{equation}
A_{\{C\}}=\frac{\sigma^+(\phi)-\sigma^-(\phi)}{\sigma^+(\phi)+\sigma^-(\phi)}
\label{eq:BCA}
\end{equation}

\item correlated charge and beam-spin asymmetries~\cite{zeiler}:
\begin{eqnarray}
&&A_{\{LU,DVCS\}}=\frac{(\sigma^+_+(\phi)-\sigma^+_-(\phi))+(\sigma^-_+(\phi)-\sigma^-_-(\phi))}
{\sigma^+_+(\phi)+\sigma^+_-(\phi)+\sigma^-_+(\phi)+\sigma^-_-(\phi)}\nonumber\\
&&A_{\{LU,I\}}=\frac{(\sigma^+_+(\phi)-\sigma^+_-(\phi))-(\sigma^-_+(\phi)-\sigma^-_-(\phi))}
{\sigma^+_+(\phi)+\sigma^+_-(\phi)+\sigma^-_+(\phi)+\sigma^-_-(\phi)}
\label{eq:BSA}
\end{eqnarray}

\item correlated charge and transversely polarized target-spin asymmetries~\cite{ave}~:
\begin{eqnarray}
&&A_{\{UT,DVCS\}}=\frac{(\sigma^+_+(\phi)-\sigma^+_-(\phi))+(\sigma^-_+(\phi)-\sigma^-_-(\phi))}
{\sigma^+_+(\phi)+\sigma^+_-(\phi)+\sigma^-_+(\phi)+\sigma^-_-(\phi)}\nonumber\\
&&A_{\{UT,I\}}=\frac{(\sigma^+_+(\phi)-\sigma^+_-(\phi))-(\sigma^-_+(\phi)-\sigma^-_-(\phi))}
{\sigma^+_+(\phi)+\sigma^+_-(\phi)+\sigma^-_+(\phi)+\sigma^-_-(\phi)}
\label{eq:TSA}
\end{eqnarray}

\end{itemize}

where the first index of the asymmetry $A$ refers to the polarization
of the beam (``U" for unpolarized and ``L" for longitudinally polarized) and the second one 
to the polarization of the target (``U" for unpolarized and ``T" for a
transversely polarized target). For $A_{UT}$, one can actually define two independent
polarization observables $A_{Ux}$ and $A_{Uy}$ corresponding to the two directions 
orthogonal to the virtual photon direction~: ``x" being in the hadronic plane and
``y" perpendicular to it (see fig.~\ref{fig:frame}).
In eqs.(\ref{eq:BCA},\ref{eq:BSA},\ref{eq:TSA}), $\sigma$ refers to the $ep\to ep\gamma$ 
cross section, the superscript being the charge of the beam and the subscript the (beam or 
target, accordingly) spin projection. 

The HERMES collaboration has measured several $\phi$ moments of these asymmetries,
leading to seventeen independent observables in all~:
\begin{eqnarray}
&&\boldsymbol{A_{\{C\}}}, A_{\{C\}}^{\sin\phi},\boldsymbol{A_{\{C\}}^{\cos\phi}}, 
A_{\{C\}}^{\cos2\phi}, A_{\{C\}}^{\cos3\phi}\nonumber\\
&&A_{\{LU,DVCS\}}, A_{\{LU,DVCS\}}^{\sin\phi}, A_{\{LU,DVCS\}}^{\cos\phi}, A_{\{LU,DVCS\}}^{\sin2\phi}\nonumber\\
&&A_{\{LU,I\}}, \boldsymbol{A_{\{LU,I\}}^{\sin\phi}}, A_{\{LU,I\}}^{\cos\phi}, A_{\{LU,I\}}^{\sin2\phi}\nonumber\\
&&\boldsymbol{A_{\{Ux,I\}}^{\sin\phi}},\nonumber\\
&&A_{\{Uy,DVCS\}},\nonumber\\
&&\boldsymbol{A_{\{Uy,I\}}} \;\;\;\;\text{and}\;\;\;\; 
\boldsymbol{A_{\{Uy,I\}}^{\cos\phi}}
\label{obs}
\end{eqnarray}

where $A^{\sin\phi}$ and $A^{\cos\phi}$ are the 
Fourier coefficients of the asymmetry and when 
no subscript is present to its simple integral normalized by $2\pi$.
In eq.~\ref{obs}, the moments written in a \textbf{bold} font are
those which are expected to be significantly different from zero
at leading twist DVCS.

\begin{table*}[h]	
\begin{tabular}{|c|c|c||c|c|c|}
\hline	
$<x_B>$~\cite{ave}  & $<Q^2>$ (GeV$^2$)~\cite{ave} & $<-t>$ (GeV$^2$)~\cite{ave}
&$<x_B>$~\cite{zeiler}  & $<Q^2>$ (GeV$^2$)~\cite{zeiler} & $<-t>$ (GeV$^2$)~\cite{zeiler} \\
\hline \hline
0.08  & 1.9 & 0.03 & 0.07  & 1.71 & 0.02 \\
0.10  & 2.5 & 0.10 & 0.10  & 2.44 & 0.08 \\
0.11  & 2.9 & 0.20 & 0.11  & 2.72 & 0.14 \\
0.12  & 3.5 & 0.42 & 0.12  & 3.63 & 0.46 \\
\hline
\end{tabular}	
\caption{The four HERMES kinematic points considered in this work. Ref.~\cite{zeiler}
has measured (correlated) beam charge and beam spin asymmetries while ref.~\cite{ave} has 
measured (correlated) beam charge and spin transverse target asymmetries. The two sets of kinematics
do not match each other perfectly but are reasonably close. In order to fit all seventeen
observables simultaneously, we have chosen to adjust all observables to the kinematics
of ref.~\cite{ave}, meaning that the beam spin asymmetries of ref.~\cite{ave} have been
slightly offset in our work.} 
\label{tab:kin}
\end{table*}	

In our study, we have considered the simultaneous fit of all these seventeen 
asymmetries. Even though many of these asymmetries are zero in the DVCS
leading twist approximation and therefore bring no constrain in our minimization
procedure, we have nevertheless decided to keep them in our fit
because they do carry some statistical significance. The fit was carried out at the 
four kinematic points displayed in the three leftmost columns of 
Tab.~\ref{tab:kin}. We show in Tab.~\ref{tab:kin} two sets of kinematics because 
the asymmetries of eq.~\ref{obs} have been obtained via two distinct 
analysis~\cite{ave,zeiler} and the two sets of kinematics do not match perfectly.  
Ref.~\cite{ave} has measured the $A_{\{C\}}$ and $A_{\{UT\}}$ 
asymmetries at the $<x_B>$,$<Q^2>$,$<-t>$ values of the first three 
(leftmost) columns of Tab.~\ref{tab:kin} while ref.~\cite{zeiler} has measured 
the $A_{\{C\}}$ and $A_{\{LU\}}$ asymmetries at the $<x_B>$,$<Q^2>$,$<-t>$
values of the last (rightmost) three columns of Tab.~\ref{tab:kin}. Actually, 
for this latter analysis, two more kinematic points are available. However,
they do not have any approximate equivalent in ref.~\cite{ave} and we have therefore  
not considered these extra data in our fit since not enough observables to fit would 
have been available. 

As the two $<x_B>$,$<Q^2>$,$<-t>$ sets approximatively match each other and, 
in order to be able to fit simultaneously all seventeen observables, we have decided to fit 
the $A_{\{C\}}$ and $A_{\{UT\}}$ asymmetries of ref.~\cite{ave} and
the $A_{\{LU\}}$ asymmetries of ref.~\cite{zeiler} at the \underline{common} 
kinematics of ref.~\cite{ave} ($i.e.$ the three rightmost columns of Tab.~\ref{tab:kin}). This
clearly introduces a slight bias for the $A_{\{LU\}}$ asymmetries, which are therefore
not calculated at the exact kinematics at which they were measured. Given the
uncertainties on our final results, which we will present shortly, this 
effect is considered negligible.

For the actual fit of the data, we have used the code of ref.~\cite{fitmick} as well 
as another independent recent code developped by H. Moutarde, based on 
analytical formulas for the $ep\to ep\gamma$ amplitude~\cite{guichonvdh} ($i.e.$ sum of BH and DVCS). 
An analysis of the DVCS JLab Halls A and B data based on this code has recently been 
performed~\cite{rv}. It has been checked that the two codes give results in agreement, the
code of ref.~\cite{rv} being simply faster.

The parameters to be fitted are the CFFs of eq~\ref{eq:eighta}-\ref{eq:eighta} 
However, like in ref.~\cite{fitmick}, we have considered only seven CFFs,
setting $\tilde{E}_{Im}$ to zero, based on the theoretical guidance which approximates 
the $\tilde{E}$ GPD by the pion exchange in the $t$-channel whose amplitude is real. 
For the minimization procedure, we have used MINUIT and, the problem at stake being non-linear 
and the parameters being correlated, MINOS for the uncertainty
calculation on the fitted parameters~\cite{james}. The function which is minimized is~:

\begin{equation}
\chi^2=\sum_{i=1}^{n}
\frac{(A^{theo}_i-A^{exp}_i)^2}{(\delta\sigma^{exp}_i)^2}
\label{eq:chi2}
\end{equation}

\noindent where $A^{theo}$ is the theoretical leading twist asymmetry, 
$A^{exp}$ is the corresponding HERMES experimental value and $\delta\sigma_{exp}$ 
is its associated experimental error bar. We have summed quadratically the systematic and 
statistical error bars of the HERMES data.

The only model dependent input in our approach is that we have bounded the domain of variation
of the seven CFFs to be fitted. Indeed, without any bounds, the fit would 
not converge or the result would depend on the starting values of the CFFs.
We have bounded the allowed range of variation of the CFFs to five times some ``reference" 
VGG CFFs, like in ref.~\cite{fitmick}. We stress that this is the only place in our work 
where some model dependency enters. However this bounded domain appears to us as a rather 
conservative choice.
We recall that the normalization of the GPDs is in general constrained by several
relations, in particular forward parton distribution functions and form factor sum rules.
Obviously, the VGG GPDs obey these relations. Furthermore, GPDs must obey dispersion 
relations~\cite{dis1,dis2,dis3,dis4} which state that the ``Re" CFFs can be 
deduced from an integral over $x$ of the ``Im" CFFs~\cite{marcmax}, so that they are not 
completely independent. Since refs.~\cite{franck,fitmick,fx,rv} seem to establish that
the VGG value of $H_{Im}$ is very reasonable, at least in the JLab regime, this
means that the VGG $H_{Re}$ should not be very much off either. However, it is to be noted 
that, in the dispersion relation formalism, for the $H$ and $E$ GPDs, there is an unknown subtraction 
constant, which can be associated to the so-called D-term~\cite{weiss} and which affects
the normalization of $H_{Re}$ and $E_{Re}$. The normalization of the D-term is
not well established but model dependent estimations~\cite{weiss} or first lattice
calculations do not seem to imply values which would escape the domain of variation
considered in the present work. 
As a last point, since the D-term is non-zero only in the $(-\xi,\xi)$ 
domain, its weight drops with $x_B$ decreasing and its influence should be significantly
reduced at HERMES energies. The influence on our results of the bounds of the domain of 
variation of the CFFs is studied furtherdown and is found, within reasonable limits,
to be modest.

\begin{figure}[htb]
\epsfxsize=10.cm
\epsfysize=10.cm
\epsffile{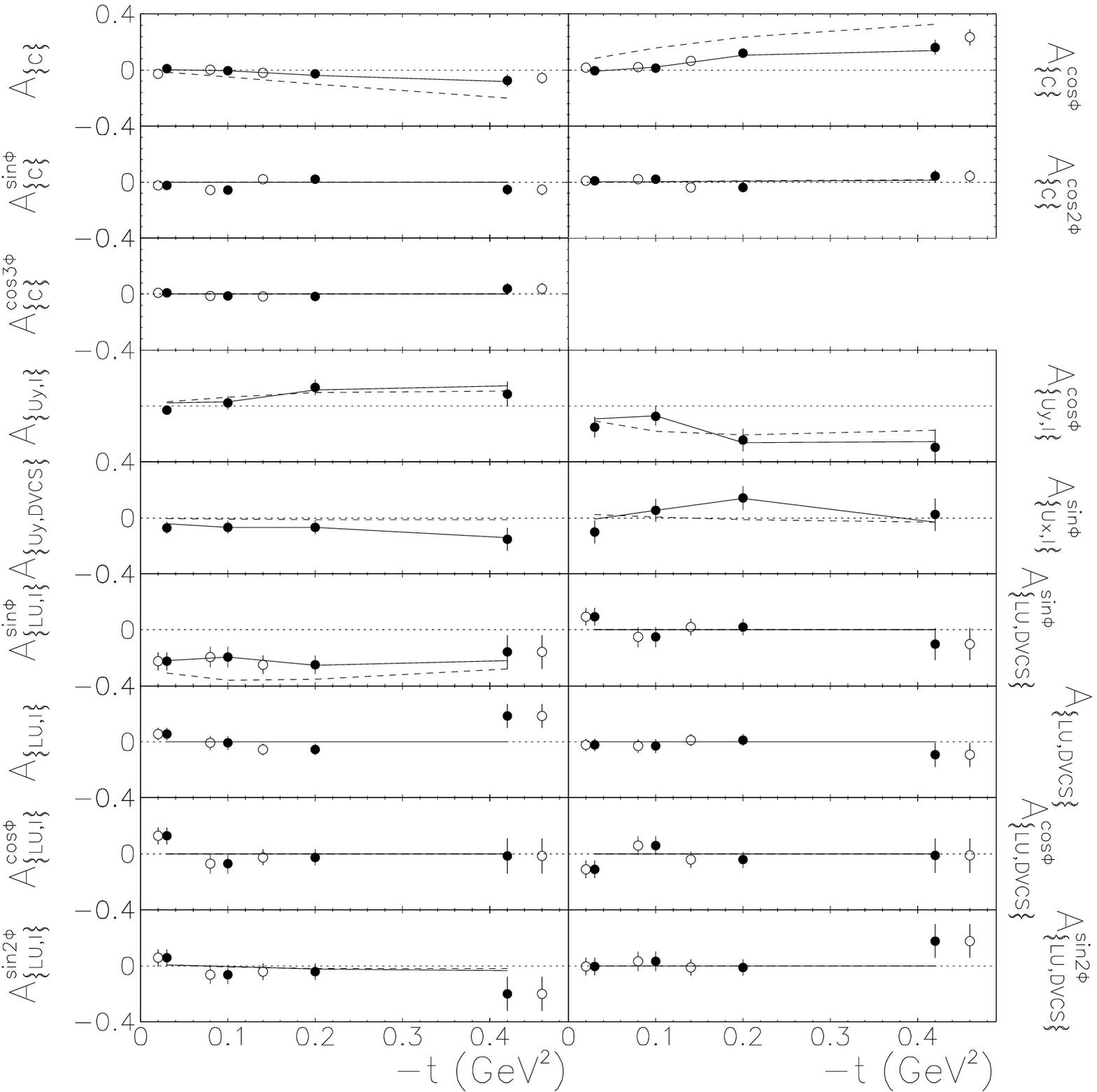}
\caption{The seventeen HERMES observables, as defined in eqs.~\ref{eq:BCA}-\ref{obs},
that have been fitted. For the first five rows, the solid circles show the HERMES data 
of ref.~\cite{ave} and the open circles show the HERMES data of ref.~\cite{zeiler}.
For the following four rows (the $A_{LU}$ asymmetries), the open circles show the HERMES data 
of ref.~\cite{zeiler} and the solid circles show these SAME data offset to the kinematics
of ref.~\cite{ave}, so as to fit all seventeen observables simultaneously at the same
kinematics. In other words, the solid circles show the data point which have been fitted.
The results of our fit are the intersections of the solid lines with the experimental points,
the result of our fits being simply linked by straight lines for the four $t$ points.
The dashed lines show the results of the reference VGG calculation, again calculated
at the four HERMES kinematic values of Tab.~\ref{tab:kin} (three leftmost columns), 
the four $t$ points being linked on the figure by straight lines.}
\label{fig:refit}
\end{figure}

The results of our fits to the seventeen HERMES asymmetries are displayed in 
fig.~\ref{fig:refit} along with the Hermes data. The numerical values of our fitted CFFs 
with their uncertainties are displayed 
in Tab.~\ref{tab:param2}. The $\infty$ symbol means in general that the MINOS 
uncertainties were ``at limit", $i.e.$ that MINOS could not reach the
$\chi^2$+1 value, defining $\sigma$, within our 7-dimensional domain which is bounded as
we recall. The $\infty$ symbol can also mean that, although
MINOS returned a finite error on the fitted parameter, this parameter actually
reached the boundary of the domain which makes this uncertainty unreliable. In such case, 
we associate an asterisk to this value in Tab.~\ref{tab:param2}. We have checked that 
the contribution to the $\chi^2$ of those CFFs which happen to reach the boundaries
of our domain are actually at the level of the percent over their whole range of variation.
Although it is not very satisfying to have fitted parameters reaching
their limit, it simply shows that these have actually little influence on our results and
that the fit is barely sensitive to them.

As can be seen from Tab.~\ref{tab:param2}, the 
five CFFs $E_{Re}$, $\tilde{H}_{Re}$, $\tilde{E}_{Re}$, $E_{Im}$ and 
$\tilde{H}_{Im}$ can in general not be constrained (except for upper or lower limits
in some cases) and they can take a whole range of values and accomodate the HERMES data.
We see that actually only the $H_{Im}$ and $H_{Re}$ CFFs come out of our fitting procedure 
with meaningful lower and upper error bars: this means that only a limited range of values 
for the two $H_{Im}$ and $H_{Re}$ CFFs is possible to fit the HERMES data. 
In other words, the HERMES data constrain only those two CFFs. This was also the 
case when we fitted the polarized and unpolarized DVCS cross sections of the 
JLab Hall A collaboration~\cite{fitmick}. In the present case, we note that the 
uncertainties on $H_{Im}$ range from $\approx$ 30\% to $\approx$50\%, which is actually less
than what we obtained for the JLab Hall A data where the error bars
were of the order of 100\%~\cite{fitmick}. This shows that,
even without normalized data such as cross sections, $i.e.$ only with a (large) series 
of asymmetries, it is possible to constrain relatively strongly $H_{Im}$ and, to a lesser extent, 
$H_{Re}$.

\begin{figure}[htb]
\epsfxsize=10.cm
\epsfysize=10.cm
\epsffile{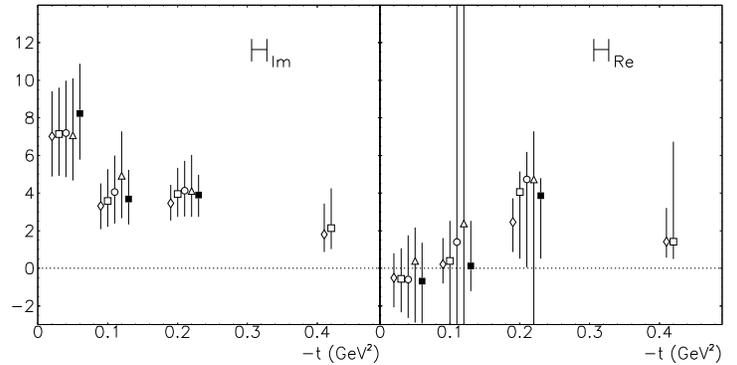}
\vspace{-4.cm}
\caption{Fitted central values and uncertainties for the $H_{Im}$ and $H_{Re}$ CFFs, fitted to
the HERMES data of fig.~\ref{fig:refit}. The open upright triangles are the results for the
domain of variation $\pm$3 times the VGG reference values, the open squares for $\pm$5,
the open circles for for $\pm$7 and the open upright triangles for $\pm$10. For these two latter
domains, there are no values displayed for the largest $t$ value as the fit did not converge
($\infty$ incertainty). The solid squares
are the results of the fit for the ``average" kinematics, $i.e.$ fixed $x_B$=0.09 
and $Q_2$=2.5 GeV$^2$, for the four $t$ values of Tab~\ref{tab:kin} -0.03, -0.10, -0.20 and -0.42 GeV$^2$
(third column). All points on this figure actually correspond to these four $t$ values:
the open upright triangles, the open circles, the open upright triangles and the solid
squares have been slightly shifted for visibility. We recall
that our definition of the ``Im" and ``Re" CFFs differ respectively of a $\pi$ factor and of
a ``-" sign with respect to the original CFF definition of ref.~\cite{kirch}.}
\label{fig:test}
\end{figure}

We have carried out several checks to ensure the reliability of our results.
First, we have studied the influence of the bounds of the domain over which the CFFs
are allowed to vary. Fig.~\ref{fig:test} shows for the $H_{Im}$ 
and $H_{Re}$ which, according to Tab.~\ref{tab:param2} are the only two CFFs with finite
uncertainties issued from our fits to the HERMES data, their fitted central values and 
uncertainties for different bounds~: $\pm$3, $\pm$5, $\pm$7 and $\pm$10 times
the VGG reference values. The figure shows that, for $H_{Im}$, at the three
lowest $t$ values, our results are essentially independent of these bounds, at the few percent
level. It can be noted that the error bars increase as the domain of variation increases,
which is naturally understandable. Only for the largest $t$ value (-0.42 GeV$^2$), some non-convergence
starts to appear if the domain of variation exceeds $\pm$5 times the VGG reference values. The 
same conclusions essentially apply to $H_{Re}$. For this latter CFF, one can note however that there 
is a more pronounced effect for the second $t$ value (-0.10 GeV$^2$). This may
be due to the
fact that $H_{Re}$ takes, at this $t$ point, a value very close to zero, even suggesting
a change of sign of this CFF in this $t$ region. Being extremely small, it is understandable
that it is difficult to fit and have sensitivity to it. Although it should be rather unrealistic
to have CFFs more than five times those given by the VGG parametrization, fig.~\ref{fig:test} shows 
that our results remain stable for larger domains, at least for the three smallest $t$ values.
Given the large MINOS uncertaintities displayed in Tab.~\ref{tab:param2}, we do not 
introduce any additional uncertainty due to this effect. Fig.~\ref{fig:test} also shows 
that one way to reduce the uncertainties on the results of our fits is, besides obviously
having more numerous and more precise data to fit, to reduce the domain of variation of the 
parameters~: the smaller the domain, the smaller the final uncertainties. In the current
study, our aim is to remain as much model-independent as possible, so we do not pursue this
direction for the moment but it is clear that if theory happens to impose some limit stronger than
the ones we have presently taken, on the values that some CFFs can take, the uncertainties on the 
fit results can only diminish.

We have also made sure that the result of the fit did not depend on the starting values
of the minimization procedure: we have taken these starting values either equal 
to zero, or to the VGG reference value, or even randomly chosen within the
range of variation of the seven CFFs (equal to, we recall, for our ``reference" configuration, 
to $\pm$5 times the reference VGG value). In each case, while the resulting fitted values for 
$E_{Re}$, $\tilde{H}_{Re}$, $\tilde{E}_{Re}$, $E_{Im}$ and $\tilde{H}_{Im}$ could vary,
the results for $H_{Im}$ and $H_{Re}$ would always converge to the same values of Tab.~\ref{tab:param2}.

\begin{table*}[htb]	
\begin{center}	
\begin{tabular}{|l|c|c|c|c|c|c|c|c|c|}
\cline{3-10}
\multicolumn{2}{c|}{}
&$H_{Re}$  & $E_{Re}$ & $\tilde{H}_{Re}$ & $\tilde{E}_{Re}$ & $H_{Im}$ & $E_{Im}$ &
$\tilde{H}_{Im}$& $\chi^2/N_{dof}$\\
\hline 
\hline 
$<x_B>$=0.08& & -0.56 & 22.87$^*$ & -1.89$^*$ & -5592.14$^*$ & 7.14 & -4.82 & 3.19$^*$ & \\
\cline{2-9}
$<Q^2>$=1.9 GeV$^2$& $\sigma^-$ & 1.78 & $\infty$ & $\infty$ & $\infty$ & 2.23 & $\infty$& $\infty$ & 1.75 \\
\cline{2-9}
$<-t>$=0.03 GeV$^2$& $\sigma^+$ & 1.62 & $\infty$ & $\infty$& $\infty$ & 2.47 & $\infty$& $\infty$& \\
\hline
\hline 
$<x_B>$=0.10& & 0.38 & 9.21 & -1.19$^*$ & 1825.39$^*$ & 3.58 & 3.37 & -1.60 & \\
\cline{2-9}
$<Q^2>$=2.5 GeV$^2$&$\sigma^-$ & $\infty$ & $\infty$ & $\infty$ & $\infty$ & 1.37 & $\infty$ & 2.92 & 1.06 \\
\cline{2-9}
$<-t>$=0.10 GeV$^2$&$\sigma^+$ & 2.11 & $\infty$ & $\infty$ & $\infty$ & 1.68 & 4.64 & $\infty$ & \\
\hline
\hline 
$<x_B>$=0.11& & 4.06 & -12.09$^*$ & -0.84$^*$ & 708.59 & 3.95 & -1.85 & -2.48$^*$ & \\
\cline{2-9}
$<Q^2>$=2.9 GeV$^2$&$\sigma^-$ & 3.53 & $\infty$ & $\infty$ & $\infty$ & 1.21 & $\infty$ & $\infty$ & 0.76 \\
\cline{2-9}
$<-t>$=0.20 GeV$^2$&$\sigma^+$ & 1.09 & $\infty$ & $\infty$ & 904.13 & 1.38 & 2.66 & $\infty$ & \\
\hline
\hline 
$<x_B>$=0.12& & 1.41 & 7.34$^*$ & 0.51$^*$ & 420.97$^*$ & 2.14 & -1.47 & 0.59 & \\
\cline{2-9}
$<Q^2>$=3.5 GeV$^2$&$\sigma^-$  & 0.91 & $\infty$ & $\infty$ & $\infty$ & 1.12 & 2.05 & $\infty$ & 1.67 \\
\cline{2-9}
$<-t>$=0.42 GeV$^2$&$\sigma^+$  & 5.32 & $\infty$ & $\infty$ & $\infty$ & 2.10 & $\infty$& 0.74 & \\
\hline
\hline 
\end{tabular}	
\caption{The CFFs and their MINOS uncertainties (negative: $\sigma^-$ and positive: $\sigma^+$)
resulting from our fit to the seventeen HERMES observables. The $\infty$ symbol means that 
the uncertainty could not be defined by MINOS, $i.e.$ that the $\chi$+1 value, defining one $\sigma$,
was out
of the variation range of our seven parameters. The corresponding fitted central value are therefore 
not constrained and meaningful. These values have been obtained with the bounds {-5,5} times 
the VGG reference values. The asterisk symbol $^*$ means that the parameter actually reached
the boundary of the domain.} 
\label{tab:param2}
\end{center}	
\end{table*}	

\begin{figure}[htb]
\epsfxsize=8.cm
\epsfysize=8.cm
\epsffile{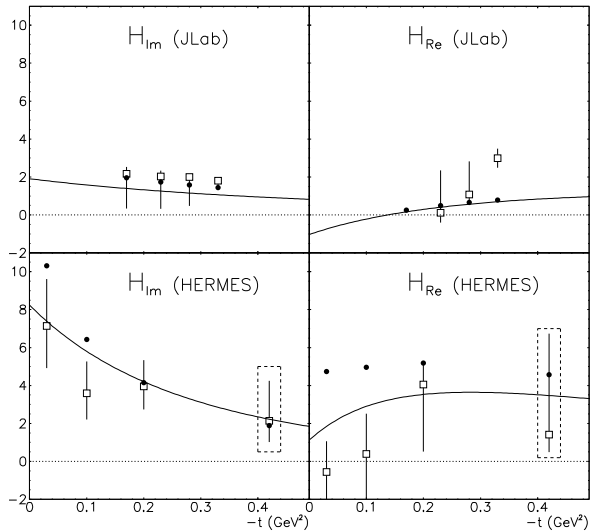}
\caption{The $t$-dependence of the $H_{Im}$ and $H_{Re}$ CFFs resulting from our fitting procedure
(open squares).
Upper panels: JLab kinematics (fit of the unpolarized DVCS cross section and of the DVCS 
beam-polarized difference of cross sections~\cite{fitmick,franck}), lower panels: 
HERMES kinematics (this work, with the values of Tab.~\ref{tab:param2}). The largest $t$ point
for the HERMES fits is in a ``box" to recall that this result is particularly dependent on the boundary
values of the domain over which the CFFs are allowed to vary (see fig.~\ref{fig:test}). 
The solid circles show the result of the reference VGG parametrisation. The solid curves show the
results of the model-based fit of ref.~\cite{fitmuller}, based on the ``high-energy" data
of HERMES~\cite{ave,zeiler}, ZEUS~\cite{zeus}, H1\cite{h1a,h1b,schoeffel} and CLAS~\cite{fx}.}
\label{fig:gpds}
\end{figure}

We also show on this figure
the prediction of the VGG model which calculate these observables from
GPD parametrizations with standard parameters. We observe
that the basic VGG paramatrisation of the GPDs already gives 
a decent description and the right trend of the data as this was also 
shown in ref.~\cite{ave}. The biggest discrepancies between the standard VGG 
calculation and the HERMES data lie in the $A_{\{C\}}^{\cos\phi}$, $A_{\{Uy,DVCS\}}$ 
and $A_{\{Ux,I\}}^{\sin\phi}$ observables.

We display on fig.~\ref{fig:gpds}, as a function of $-t$, the resulting $H_{Im}$ and 
$H_{Re}$ CFFs issued from our fit to the HERMES data (lower panels) along with the 
same two CFFs issued from the fit of the JLab Hall A data (upper panels). 
The values and kinematics of these two latter ``JLab" CFFs are recalled in Tab.~\ref{tab:jlab}.
The kinematics of the two experiments do not perfectly match~: for instance $Q^2$=2.3 GeV$^2$ 
for the JLab Hall A data while $Q^2$ ranges from 1.9 to 3.5 GeV$^2$ for HERMES 
(for an average $<Q^2>$=2.5 GeV$^2$~\cite{ave}). However, in the
leading-twist handbag and leading order QCD approximation, which is the frame in which
this work is done, we do not take into account any $Q^2$ dependence of the CFFs. Such $Q^2$ 
dependence can actually not be really resolved with the present available data sets.
Also, while the JLab data are at a fixed $x_B$ (=0.36)
the HERMES data have been taken at varying $x_B$ (from 0.07 to 0.12, resulting
in an average $<x_B>$ of 0.09~\cite{ave}). However, the solid square points on fig.~\ref{fig:test}
show that taking these average HERMES $<x_B>$ and $<Q^2>$ affect only at the percent level 
the $H_{Im}$ and $H_{Re}$ values of Tab.~\ref{tab:param2}.
We also show on fig.~\ref{fig:gpds}, the standard VGG values for these two CFFs. 

Although the error bars on the fitted CFFs are quite significant, some general features and 
trends can nevertheless already be distinguished on fig.~\ref{fig:gpds}~:
\begin{itemize}
\item At fixed $-t$, $H_{Im}$ takes higher values at HERMES 
than at JLab kinematics (for instance, at $-t\approx$ 0.2 GeV$^2$, 
$H_{Im}\approx$ 2 at JLab (see the asymmetric uncertainties in Tab.~\ref{tab:jlab}) 
while $H_{Im}\approx$ 4 at HERMES (see the asymmetric uncertainties in Tab.~\ref{tab:param2}). 
This means that $H_{Im}$ rises with decreasing $\xi$~: $\xi$=0.22 ($x_B$=0.36)
at JLab while $\xi$=0.06 ($x_B$=0.11) at HERMES (for $-t$= 0.2 GeV$^2$).
This is reminiscent of the $x$-dependence of the standard proton unpolarized parton
distribution as measured in DIS, to which $H_{Im}$ reduces in forward
kinematics ($\xi=t=0$).
\item Comparing at small $t$ the values of $H_{Im}$ and $H_{Re}$, one sees
that $H_{Im}$ is largely dominant ($H_{Re}$ is in fact compatible with zero
within one sigma for the smallest $t$ values). This is reminiscent 
of the Regge prediction that diffractive processes have a
dominant imaginary amplitude at high energies (see ref.\cite{IntroRegge} for instance). 
\item At both energies, $H_{Im}$ decreases with $-t$ while $H_{Re}$
increases (at least up to $-t\approx$ 0.3 GeV$^2$) with, for this latter
CFF, possibly a change of sign, considering the central values of the fit,
starting negative at small $-t$ and reaching positive 
values at larger $-t$. Concerning $H_{Im}$, the $t-$slope appears to increase
with the energy rising.
\item As a side point, the VGG prediction for $H_{Im}$ is in relative good agreement with
the fitted value, at the 15\% level in general, at most 30\%, for both JLab
and HERMES kinematics. For $H_{Re}$, the comparison with the VGG parametrisation 
is much worse, the disagreement being the strongest, at HERMES kinematics, 
for the lowest $t$ value.
As mentionned earlier, the D-term whose normalization is barely known, could be
an explanation for this. We recall that, at a couple of instances, it has been alluded 
or speculated that 
an important piece in the VGG GPD parametrization contributing to the real part of the
handbag amplitude, could be missing~\cite{dual,photon07,rhoclas}. 
\end{itemize} 

\begin{table*}[h]	
\begin{tabular}{|c|c|c|c|c|c|}
\hline	
$<-t>$ (GeV$^2$) & -0.17 & -0.23 & -0.28 & -0.33 \\
\hline \hline
$H_{Im}$ & 2.17 & 2.03 & 2.00 & 1.80\\
$\sigma^-$ & 1.84 & 1.70 & 1.53 & 0.40 \\
$\sigma^+$ & 0.35 & 0.29 & 0.15 & 0.06 \\
\hline \hline
$H_{Re}$ &   & 0.11 & 1.08 & 2.99 \\
$\sigma^-$ & & 0.51 & 0.53 & 0.50\\
$\sigma^+$ & & 2.24 & 1.74 & 0.49\\
\hline
\end{tabular}	
\caption{Reminder of the values and their uncertainties of the $H_{Im}$ and $H_{Re}$ CFFs
found in ref.~\cite{fitmick}, resulting from the fit of the JLab Hall A unpolarized 
DVCS cross section and DVCS beam-polarized difference of cross sections~\cite{franck}.} 
\label{tab:jlab}
\end{table*}	

Finally, the solid curves on fig.~\ref{fig:gpds} show the result of the model-based fit
of ref.~\cite{fitmuller}, which we will call K-M in the following. In this work, only the $H$ GPD, 
which is parametrized by a Regge-inspired functional form, is considered. This model is well suited 
to describe the low $x_B$ regime and its parameters are fitted to the DVCS data 
of HERMES~\cite{ave,zeiler}, ZEUS~\cite{zeus}, H1~\cite{h1a,h1b,schoeffel} and,
additionnally, of CLAS~\cite{fx} for the low energy part. 
As can be seen from fig.~\ref{fig:gpds}, it is remarkable that the corresponding 
K-M $H_{Im}$ and $H_{Re}$ CFFs have values and behaviors as a function of $t$ in relatively 
good agreement with ours at both HERMES and JLab energies. The two studies 
being independent and based on relatively different approaches, this gives good confidence 
in the present results. The only important discrepancy lies in the JLab large $t$ $H_{Re}$ CFF 
for which we obtain a higher value. The inclusion of the JLab Hall A DVCS cross sections in 
the K-M fit, which could shed light on this issue, requires extra assumptions and 
complications beyond the simple dominance of $H$ (the introduction of $\tilde{H}$ in particular) 
and we therefore do not dwell on this point, which is beyond the scope of this article.

In summary, we have fitted for the first time in a model independent way (at the QCD leading 
twist and leading order), the recently published HERMES DVCS data, consisting of seventeen (correlated) 
beam charge, beam spin and target transverse spin target asymmetries. We have been able to extract 
numerical values for the $H_{Im}$ and $H_{Re}$ CFFs with finite uncertainties (in particular,
of the order of 30\% for $H_{Im}$). We recall that these uncertainties arise from
the extremely conservative approach that we have decided to adopt as a first step, of bounding the 
domain of variation of the fitted CFFs by a factor 5 with respect to the VGG CFFs reference values. 
It is remarkable that with such loose constrain already finite results come out. This encourages to 
pursue such approach with the extra input of bounding more severely the domain of variation allowed 
for the CFFs, based on as model independent as possible theoretical guidance. 
We leave such study for later.

Comparing the results that we obtained with the HERMES data to those obtained with the same fitting 
procedure with the JLab lower energies data~\cite{fitmick}, in spite of the important error bars, 
we have begun to hint at a few key features about the energy ($x_B$) and $t$ dependence of these two CFFs: 
in particular, the dominance and increase of $H_{Im}$ with increasing energy accompanied by the sharpening
of its $t$ slope. In contrast, $H_{Re}$ tends to exhibit a rising $t$ slope, with 
possibly hints of a sign change. We have also concluded
that the VGG parametrisation of $H_{Im}$ seems valid at the 30\% level while the VGG
parametrisation of $H_{Re}$ is most likely much less reliable.
A rich harvest of precise and numerous DVCS data and observables (cross sections, polarized beam 
and/or target, double polarization observables,...), providing sensitivity to new
CFFs other than $H_{Im}$ and $H_{Re}$ and improved precision in these latter, are expected in the 
near future from JLab and HERMES. Fitting analysis, model independent or using educated
parametrizations such as in ref.~\cite{fitmuller}, are necessary to 
fully exploit these data and further reveal the space-momentum structure of the nucleon.

We are very thankful to E. Avetysian, M. Diehl, F. Ellinghaus, D. Hasch, D. M\"uller, C. Riedl, 
G. Schnell and M. Vanderhaeghen for very rich and useful discussions and to the HERMES collaboration 
for providing tables of their data (and the invitation to M. G. for a seminar on
this work). Additional and particular thanks go to D. M\"uller for providing numerical 
values and feedback about the K-M model-based fit. This work was supported in part by the 
French ``Nucleon GDR" n$^o$ 3034.  

\newpage 

\end{document}